\documentclass[conference]{IEEEtran}
\IEEEoverridecommandlockouts
\usepackage{cite}
\usepackage{amsmath,amssymb,amsfonts}
\usepackage{graphicx,color,overpic,psfrag}
\usepackage{xcolor}
\usepackage{algorithmic}
\usepackage{graphicx}
\usepackage{textcomp}
\usepackage{fancyhdr}
\usepackage{xcolor}
\usepackage{graphicx}
\usepackage{bm}
\usepackage{subfigure}
\usepackage{multirow}
\usepackage{booktabs}
\usepackage{caption}
\usepackage{booktabs}
\usepackage{amsmath}
\usepackage[numbers,sort&compress]{natbib}
\usepackage{algorithm}
\usepackage{algorithmic}
\usepackage{amsfonts}
\def\BibTeX{{\rm B\kern-.05em{\sc i\kern-.025em b}\kern-.08em
    T\kern-.1667em\lower.7ex\hbox{E}\kern-.125emX}}

\newcommand{\bl}[1]{\color{black}#1}
    
\begin{document}

\title{Keypoint Detection Empowered Near-Field User Localization and Channel Reconstruction \\

\author{Mengyuan Li, \textit{Student Member, IEEE}, Yu Han, \textit{Member, IEEE}, Zhizheng Lu, \textit{Student Member, IEEE}, \\
Shi Jin, \textit{Fellow, IEEE}, Yongxu Zhu, \textit{Senior Member, IEEE}, and Chao-Kai Wen, \textit{Fellow, IEEE}\\
}
\thanks{M. Li, Y. Han, Z. Lu, S. Jin, and Y. Zhu are with the National Mobile Communications Research Laboratory, Southeast University, Nanjing 210096, China (email: mengyuan$\_$li@seu.edu.cn; hanyu@seu.edu.cn; luzz@seu.edu.cn; jinshi@seu.edu.cn; yongxu.zhu@seu.edu.cn).

C.-K. Wen is with the Institute of Communications Engineering, National Sun Yat-sen University, Kaohsiung 804, Taiwan (e-mail: chaokai.wen@mail.nsysu.edu.tw).

This is an extended and revised version of a previous conference paper that was presented in  IEEE 99th Veh. Technol. Conf. (VTC-Spring) \cite{VTCkeypoint}.
}
}
\maketitle

\begin{abstract}
 
In the near-field region of an extremely large-scale multiple-input multiple-output (XL MIMO) system, channel reconstruction is typically addressed through sparse parameter estimation based on compressed sensing (CS) algorithms after converting the received pilot signals into the transformed domain. However, the exhaustive search on the codebook in CS algorithms consumes significant computational resources and running time, particularly when a large number of antennas are equipped at the base station (BS). To overcome this challenge, we propose a novel scheme to replace the high-cost exhaustive search procedure. We visualize the sparse channel matrix in the transformed domain as a channel image and design the channel keypoint detection network (CKNet) to locate the user and scatterers in high speed. Subsequently, we use a small-scale newtonized orthogonal matching pursuit (NOMP) based refiner to further enhance the precision. Our method is applicable to both the Cartesian domain and the Polar domain. Additionally, to deal with scenarios with a flexible number of propagation paths, we further design FlexibleCKNet to predict both locations and confidence scores. Our experimental results validate that the CKNet and FlexibleCKNet-empowered channel reconstruction scheme can significantly reduce the computational complexity while maintaining high accuracy in both user and scatterer localization and channel reconstruction tasks.

\end{abstract}


\section{introduction}
Massive multiple-input multiple-output (MIMO) technology stands at the forefront of advancements in the fifth generation (5G) communication systems, providing significant gains in data transmission rate and energy efficiency \cite{1}. As both the academia and the industry look ahead to the advent of future sixth generation (6G) wireless systems, there is palpable anticipation for even greater leaps in communication performance, including a 100-fold increase in peak data rate, a 10-fold reduction in latency, and a 10-fold improvement in connection sparsity to cater to emerging applications such as virtual reality and augmented reality \cite{2, 5survey}. This anticipation underscores the critical role that extremely large-scale (XL) MIMO, with its significantly augmented number of antennas, is poised to play in meeting the escalating demands of future communication systems.
 
However, the change from massive MIMO to XL MIMO transcends mere increases in the number of antennas. It fundamentally reshapes the characteristics of the channel, heralding a paradigm shift. This transition brings forth some new challenges, especially in migrating from the conventional far-field uniform plane wave to the new non-uniform spherical wave (NUSW) propagation \cite{219survey}. In the far-field region, the channel phases are modeled linearly and the amplitudes are modeled uniformly across the array elements. But in the near-field region, this phenomenon no longer exists. Moreover, the near-field channel model is no longer solely dependent on the angle of arrival, it also correlates with the distance from the user or the scatterer to the antenna. NUSW is more general and is required to accurately characterize both the phase and amplitude variations across the array elements. Additionally, along with the progressively shrinking cell size and the rapidly growing Rayleigh distance due to the deployment of XL MIMO at the BS, the users or scatterers are more likely to be distributed in the near-field region. The traditional assumptions of far-field propagation no longer suffice, necessitating a paradigm shift in the channel estimation problem to obtain the channel state information (CSI), which serves as a guiding factor for transceiver design and other applications. In the near-field region, the spatial resolution of the channel becomes paramount, requiring tailored channel estimation techniques that can accurately capture the spatial variations and multi-path effects inherent in this environment. Moreover, with the expansion scale of the antenna array, the dimension of the channel matrix increases dramatically, leading to a further explosion in the computational complexity of channel estimation, resulting in increased communication latency and computational overhead. To tackle these challenges and develop efficient channel estimation schemes for XL MIMO systems, researchers have been focusing on the following approaches for several years, including statistical characteristics-based channel estimation \cite{20survey, 207survey}, sparsity-based channel estimation \cite{cs2, cs3, cs1, 5, 7,8}, and machine learning-based channel estimation \cite{206survey, 6, mrdn, 219survey, dl1, dl2, dl3, dl4}.

The first approach is to model the channel based on statistical characteristics such as the channel correlation matrix. \cite{20survey, 207survey} used minimum mean square error (MMSE) channel estimation to ensure a low normalized mean-square error (NMSE). However, the prerequisites of knowing the complete knowledge of the spatial correlation matrix are difficult to meet due to its extremely high dimension. Additionally, the computational complexity of MMSE channel estimation in XL MIMO systems is very high. An alternative approach, which is relatively efficient though less accurate, is based on least squares (LS) estimation. It does not need to know the complete prior knowledge of the channel and can achieve satisfactory NMSE performance. For this approach, how to balance the performance with the efficiency is still a key issue.

The second approach is based on exploiting the latent sparsity of the channel in transformed domains and using CS algorithms to estimate the sparse parameters \cite{cs1, cs2, cs3}. In the far-field region, the most widely applied method is to transform the original channel matrix into the angular domain showing sparsity characteristics. This can be achieved by multiplying the original matrix with a standard Fourier matrix sampled from the angular domain. \cite{cs1} used the classical orthogonal matching pursuit (OMP) algorithm to estimate the parameters and reconstruct the channel. \cite{5} proposed a more accurate CS-based algorithm that utilizes a newtonized refiner to further improve the performance. However, these CS-based methods have two main drawbacks. Firstly, they are not suitable for the near-field region in XL MIMO systems anymore due to the diminishing angular-domain sparsity. Secondly, using iterative algorithms such as OMP for channel estimation results in extremely high computational complexity, especially in scenarios with a large number of antennas. To solve the first problem, \cite{7,8} proposed the Cartesian-domain channel representation and the Polar-domain channel representation. The transform matrix in the Cartesian domain is generated by uniformly sampling in the 2D plane of the z-x coordinate system, and the transform matrix in the Polar-domain is obtained by uniformly sampling in the angular domain and non-uniformly sampling in the distance domain. The channel matrix in both Cartesian domain and Polar domain show sparsity again, which can be subsequently leveraged by the CS-based channel estimation algorithms. However, given the codebook in the Cartesian or the Polar domain, the procedure of a two-dimensional exhaustive search over the whole codebook and calculations of the projection coefficients consume much computational resources. With the increase of the number of antennas, the Rayleigh distance grows, necessitating a broader sampling range for the codebook. This leads to a larger number of codewords and a rapid escalation in the complexity of the CS-based methods. Therefore, even though these CS-based channel estimation methods can achieve high accuracy, their computational complexity is still very high in XL MIMO systems, and it is difficult to apply to real communication systems.

For the third approach, several studies have integrated machine learning (ML) techniques, adopting either data-driven approaches or dual data-model-driven methodologies to reduce computational complexity. For example, \cite{6} employs an object detection network to replace the exhaustive search on the angular-domain codebook in massive MIMO systems for the far-field region channel estimation. Through such neural networks, all path parameters can be extracted in a single-round inference, obviating the need for exhaustive searches on the codebook and greatly reducing the computational complexity. The subsequent newtonized optimizer can further improve estimation accuracy and make it comparable to NOMP algorithm \cite{nomp}. Moreover, some studies have explored the use of denoising neural networks. The multiple residual dense network (MRDN) was proposed in \cite{mrdn} by exploiting the angular-domain channel sparsity, estimating the distribution of the noise, and removing it from the received noisy signal. \cite{219survey} is based on MRDN and further designed the Polar-domain MRDN (PMRDN) with an atrous spatial pyramid pooling-based residual dense network (ASPP-RDN) and improved the estimation accuracy. It transmits the received signals into the Polar domain and estimates the original channel matrix in the Polar domain, recovering the original channel matrix through inverse Polar transformation. The performance of the denoising-based methods surpasses OMP, but the complexity is approximately twice that of OMP. Although not as abundant, there are several research efforts that explore the use of ML to address near-field channel estimation problems and have achieved promising results \cite{dl1, dl2, dl3, dl4}. How to balance the computational complexity and the channel estimation accuracy remains a crucial task worthy of further exploration.

% Commnet: 注意， "所有" 圖的 label, legend 都太小了
% 審稿人反映了這問題，但所有的圖仍都沒修改

\begin{figure*}[t]
\centering
\includegraphics[height=6.08519cm,width=18cm]{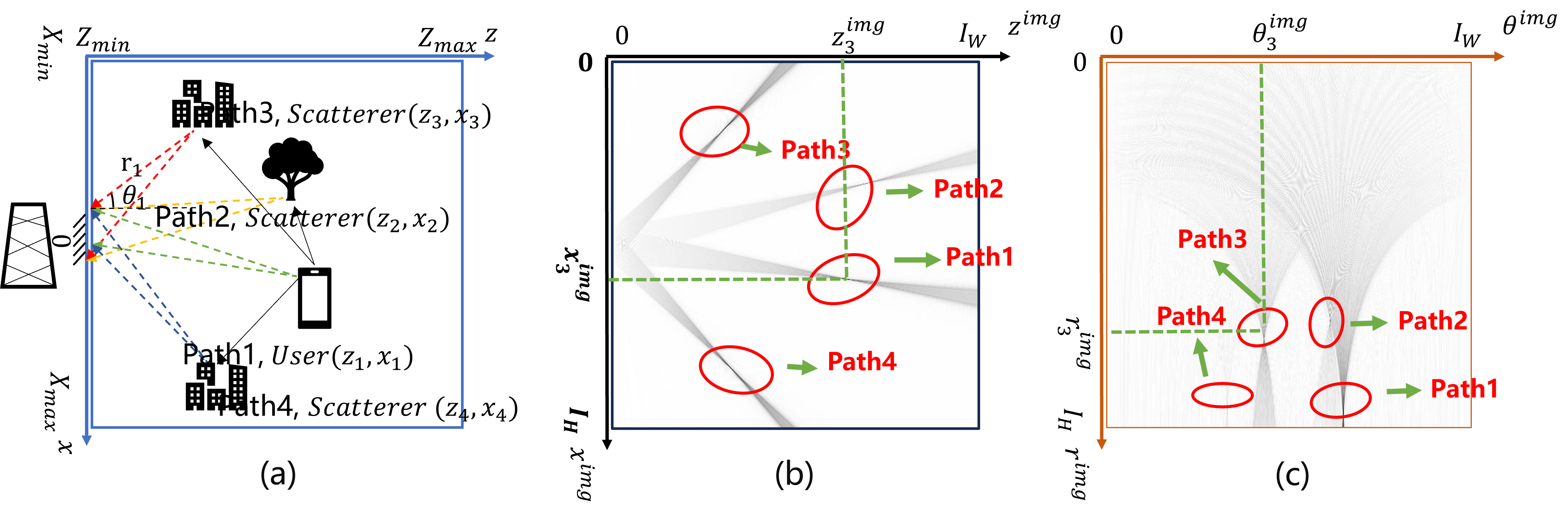}
\caption{(a) An example of near-field channel model in the real communication system, and 1 user and 3 scatterers are located in the observed region. (b) An example of near-field channel image in the Cartesian domain, and there are 4 intersecting X-shaped energy convergence zones. (c) An example of near-field channel image of the Polar domain, and there are 4 intersecting hourglass-shaped energy convergence zones.} 
\label{channelimages}
\end{figure*}

In this paper, we propose a novel approach to address the challenges of near-field channel estimation in XL MIMO systems. Our method leverages recent advancements in deep learning and sparse signal processing to formulate the parameter estimation problem as a keypoint detection task in sparse channel images. This approach provides a comprehensive solution that can achieve both high computational efficiency and estimation accuracy. It mainly consists of two main stages: coarse parameter estimation and parameter refinement. In the coarse parameter estimation phase, we design CKNet to locate the user and scatterers within the observed region through a single-round network inference. Then, we employ a small-scale NOMP refiner to further enhance the accuracy. This two-stage channel estimation scheme has high accuracy with low computation cost. Furthermore, it is applicable to both the Cartesian domain and the Polar domain. Additionally, to adapt this method to the scenario where the number of paths is flexible, we further design the FlexibleCKNet, a variation of CKNet. To evaluate the localization and channel reconstruction performance of our proposed algorithm, we conduct extensive simulations to measure the L1 Distance and the NMSE. Our results demonstrate that this keypoint detection-empowered approach achieves high accuracy in both channel reconstruction and user localization tasks under different scenarios with a wide range of signal-to-noise ratios (SNR). At the same time, compared with the high-precision near-field newtonized orthogonal matching pursuit (NNOMP) algorithm \cite{nnomp1}, it can greatly reduce the computational complexity.

In the following section, we first introduce the system model. The mechanism of keypoint detection-based parameter estimation and the details of the proposed channel reconstruction scheme are provided in Sections III and IV, respectively. Section V evaluates the scheme, and Section VI concludes the paper. 

\textit{Notations}--We denote scalars by letters in normal fonts and use uppercase and lowercase boldface letters to represent matrices and vectors, respectively. $\mathcal{N}$ represents the Gaussian distribution. The superscripts $(\cdot)^{\top}$ and $(\cdot)^H$ indicate transpose, and conjugate transpose, respectively. $\mathbb{E}\{\cdot\}$ means considering the expectation with respect to the random variables inside the brackets. We also denote the absolute value and modulus operations by $|\cdot|$ and $\|\cdot\|$.

\section{System Model}

In a single-cell XL MIMO system, the BS is equipped with a uniform linear array (ULA) with $N$ antennas uniformly spaced at an interval of $d$, while the user equipment (UE) is equipped with a single antenna. The carrier frequency is denoted by $f_c$, and the carrier wavelength is $\lambda = {c}/{f_c}$, where $c=3 \times 10^8$ represents the speed of light. The ULA is centered at the origin of a z-x 
coordinate system, with the antennas positioned along the $x$-axis. Assuming that there are $S-1$ scatterers between the BS and the UE, and we consider only the last-jump scatterers. The complex channel matrix is represented by $\mathbf{h} \in \mathbb{C}^{N \times 1}$.

In an XL MIMO system, the considerable increase in the number of antennas significantly enlarges the array aperture to $(N-1)d$, thereby increasing the probability of users and scatterers falling within the near-field region. {\bl As the near-field effect becomes prominent, it is important to take amplitude and phase deviations across the array into account since the distances from different antennas to the user or the scatterer can no longer be considered identical}. The Rayleigh distance is used to distinguish between the near-field and far-field regions, which is defined as $d_{\mathrm{R}}= {2((N-1) d)^2}/{\lambda}$. When the distance from the user or scatterer to the array is less than the Rayleigh distance, they fall into the near-field region, where amplitude and phase deviations across the array become significant. In an upper mid-band scenario featuring $N=1024$ antennas and $f_c = 6$ GHz, the Rayleigh distance $d_{\mathrm{R}}$ exceeds 26 km, significantly surpasses the typical single-cell radius. In such cases, a more precise channel model tailored for the near-field region becomes imperative for accurate system characterization and performance evaluation. 

In Fig. \ref{channelimages} (a), the signal transmitted by the user may propagate directly to the BS via the line-of-sight (LoS) path or be reflected by some scatterers along non-line-of-sight (NLoS) paths. For NLoS paths, we assume the entire array serves as the observed region. The coordinates of the $s$-th scatterer are denoted as $(z_s, x_s)$, where $ z_s \in\left[Z_{\text {min }}, Z_{\text {max }}\right]$ and $ x_s \in\left[X_{\text {min }}, X_{\text {max }}\right]$, and $Z_{\text{min}}, Z_{\text{max}}, X_{\text{min}}, X_{\text{max}}$ represent the bounds of the observed region. Moreover, we use $(r, \theta)$ to stand for the positions of the user or scatterers in the Polar domain. Here, $r$ signifies the distance from the user or scatterer to the central antenna, while $\theta$ represents the angle between the line connecting the user or scatterer and the central antenna, and the perpendicular line from the antenna. The angle $\theta$ ranges between $[-\pi/2, \pi/2]$. The transformation between $[z,x]$ in the Cartesian coordinates and $[r,\theta]$ in the Polar coordinates is expressed as follows:

\begin{equation}
r=\sqrt{x^2+z^2}, ~~
\theta=\arctan \left(\frac{x}{z}\right).
\label{eq:angle_distance}
\end{equation}

% Comment: 注意，上方你用 $[z,x]$ $[r,\theta]$
% 下方 (z,x) --> $[z,x]$
% 後方都需要一致，請檢視整個文章

% Solved

{\bl
In the near-field region, due to the spherical wavefront of the wireless signal, the array response induced by a user or scatterer at position $[z,x]$ is denoted as $\mathbf{a}(z, x) \in \mathbb{C}^{N \times 1}$. Here, the $n$-th entry of the steering vector in the Cartesian domain $\mathbf{a}(z, x)$ can be expressed as:
\begin{equation}
[{\bf a} (z, x)]_n=\frac{1}{D_n(z, x)} \cdot e^{-j k_c D_n(z, x)},
\label{eq:steering_vector}
\end{equation}
where $n \in\left[(1-N)/2, (N-1)/2\right]$, $k_c= {2 \pi}/{\lambda}$, and 
\begin{equation}
D_n(z, x)=\sqrt{z^2+\left(x-n \cdot \frac{d}{2}\right)^2}
\end{equation}
represents the distance between the user or scatterer and the $n$-th antenna.
The multi-path channel matrix, i.e., $\mathbf{h} \in \mathbb{C}^{N \times 1}$, can be expressed as
\begin{equation}  
\mathbf{h}=\sum_{s=1}^S g_s \mathbf{a} (z_s, x_s),
\label{eq:channelmodel1}
\end{equation}
where $g_s$ represents the complex gain of the $s$-th path, and $(z_s, x_s)$ is the coordinate of the user or the $s$-th last-hop scatterer. Here, we use $s=1$ to represent the LoS path and $s>1$ to represent the $s$-th NLoS path. 
Similarly, the channel model can be represented by the Polar domain parameters as
Similarly, the channel model can be represented by the Polar domain parameters as
\begin{equation}  
\mathbf{h}=\sum_{s=1}^S g_s \mathbf{b} (r_s, \theta_s),
\label{eq:channelmodel2}
\end{equation}
and the $n$-th entry of the steering vector is
\begin{equation}
[\mathbf{b}(r, \theta)]_n=\frac{1}{D_n(r, \theta)} e^{j k_c D_n(r, \theta)},
\end{equation}
where the distance is expressed as
\begin{equation}
D_n(r, \theta)=\sqrt{r^2+n \cdot d \cdot r \cdot \sin \theta+\frac{n^2 d^2}{4}}.
\end{equation}
}

We can estimate the uplink channel during the uplink-sounding phase. Without loss of generality, the pilots are set as all-1 signals. Therefore, the received pilot signals at the BS, denoted as $\mathbf{y}\in \mathbb{C}^{N \times 1}$, can be represented as 
\begin{equation} \label{eq:y}
\mathbf{y}=\sqrt{P}\mathbf{h}\mathbf{x} +\mathbf{n},  
\end{equation} 
where $P$ is the average transmission power, $\mathbf{x}$ is the transmitted pilot signal, and $\mathbf{n} \in \mathbb{C}^{N \times 1}$ is the additive Gaussian complex noise, following the distribution $\mathcal{N}\left(0, \sigma^2\mathbf{I}_N\right)$ with $\mathbf{I}_N$ being the identity matrix.

\section{Acquire model parameters through keypoint detection}

Given the parametric channel model (\ref{eq:channelmodel1}) and (\ref{eq:channelmodel2}), we can reconstruct the channel utilizing the path-related parameters. Here, we take channel model in the Cartesian domain as an example, and channel model in the Polar domain is similar. The parameters include the coordinates $(z_s,x_s)$ of the $s$-th user or scatterer and the complex gain $g_s$. Our task is to estimate $\{ x_s,z_s,g_s \}$ for $s = 1,\ldots,S$ from the received signal $\mathbf{y}$.The channel reconstruction problem can be converted to a finite parameter estimation problem. The previous work \cite{nnomp1} has proposed the NNOMP algorithm, a high-precision algorithm for this issue extending NOMP algorithm to the near-field region. However, it also has some drawbacks that make it difficult to apply to real communication systems.

In this section, we formulate two key problems that lie in the NNOMP algorithm and propose an efficient parameter estimation strategy that designs a lightweight keypoint detection network, i.e., CKNet, to obtain accurate locations of all paths through a single-round network inference.

\subsection{Challenges of NNOMP algorithm}

The NNOMP algorithm comprises a new path detection phase and a cyclic refinement phase. When the residual power of the $t$-th iteration $\|\mathbf{y}_{r,\textit{t}}\|^2 >= \tau$, where $\tau$ is the threshold, an exhaustive search is conducted on the whole Polar-domain codebook to detect new path. Subsequently, $R_C$ cycles of refinement are performed to refine all the estimated parameters utilizing newtonized optimizer. While this method can reach high accuracy, the complexity is also significant, especially for the codebook search process, whose computational complexity is expressed as $\mathcal O(\hat{S}NN_{Z}'N_{X}')$. Here, $\hat{S}$ represents the number of estimated paths, and $N_{Z}'$ and $N_{X}'$ denote the codebook sizes in dimensions $Z$ and $X$, respectively. The complexity escalates rapidly with their increase, resulting in significantly prolonged channel estimation time. Therefore, we propose the first question:

{\bf Q1: How can we extract parameters of all paths with low complexity?} 
When estimating parameters using NNOMP, for each path, a complete search is conducted on the whole codebook to find out the best-suited codeword. After that, newtonized optimizer is applied to further fine-tune the estimation. The exhaustive search process consumes significant time and computational resources. Therefore, it is valuable to investigate methods that can quickly extract all path parameters with low complexity.

\subsection{Sparse channel image in the transformed domain}

As illustrated in Fig. \ref{channelimages}, we can efficiently extract parameters by converting the original channel matrix into sparse domains, including the Cartesian domain and the Polar domain, where the paths exhibit noticeable sparsity and directionality. The intersection point of each propagation path possesses the highest energy, and the locations of intersection points follow the property I. 

\textit{Property 1:} In the transformed domain channel image, the coordinates of the intersection points can be approximately considered as the positions of users or scatterers, i.e., $(x_s,z_s)$ or $(r_s, \theta_{s})$.

\textit{Proof:} Refer to Appendix A. 

Consequently, we can obtain path parameters from the transformed domain channel image, i.e., $(x_s,z_s), ~{s=1, \ldots, S}$. We can leverage neural networks to complete this task by learning features from training samples and then extracting keypoints from the tested transformed domain channel images. By crafting a lightweight neural network, the computational complexity during inference can be substantially reduced compared to exhaustive searching in NNOMP, thereby we can address question {\bf Q1} posed in section \uppercase\expandafter{\romannumeral3}.A.

We can convert the received signal $\mathbf{y}$ into the transformed domain, denoted by ${\mathbf{y}}_\mathrm{T} \in \mathbb{C}^{N_Z N_X\times 1}$. Specifically, 
\begin{equation}
{\mathbf{y}}_{\mathrm{T}}=\mathbf{U}_{\mathrm{T}} \mathbf{y},
\label{eq:transform}
\end{equation}
where $\mathbf{U}_{\mathrm{T}}$ represents the transformed matrix. We select two transformed domains for algorithm design, including the Cartesian domain and Polar domain. The transformed matrices are denoted as $\mathbf{U}_{\mathrm{C}} \in \mathbb{C}^{N_X N_Z \times N}$ and $\mathbf{U}_{\mathrm{P}} \in \mathbb{C}^{N_R N_{\Theta} \times N}$ respectively, where $N_X$ and $N_Z$ are the numbers of sampling points on $X$-axis and $Z$-axis, $N_R$ and $N_{\Theta}$ are the numbers of sampling points on $R$-axis and $\Theta$-axis, respectively. {\bl Similar to the sampling scheme in \cite{10}, we adopt uniform sampling to collect codewords for the Cartesian domain. For the Polar domain, we employ uniform sampling along the angular axis and logarithmic sampling along the distance axis.}

The sampling point $(\bar{z}, \bar{x})$ lies within the range of $[(Z_{\min },X_{\min }),(Z_{\max },X_{\max })]$, and
\begin{subequations}
\begin{align}
\bar{z}&= \left\{ Z_{\min }, \, Z_{\min }+\Delta Z, \ldots, \, Z_{\max } \right\}, \\
\bar{x}&= \left\{ X_{\min }, \, X_{\min }+\Delta X, \ldots, \, X_{\max } \right\},
\end{align}
\end{subequations}
where $\Delta Z$ and $\Delta X$ are the sampling intervals on the $z$ axis and $x$ axis, respectively, and
\begin{equation}
\Delta Z=\frac{Z_{\text {max }}-Z_{\text {min }}}{N_Z},  ~~\Delta X=\frac{X_{\text {max }}-X_{\text {min }}}{N_X}.
\end{equation}
For the Polar domain, the pre-defined region lies within the range of $[(R_{\min },\Theta_{\min }),(R_{\max },\Theta_{\max })]$. And $(\bar{r}, \bar{\theta})$ are the uniform sampling point and logarithmic sampling point, respectively:
\begin{subequations}  
\begin{align}
\bar{\theta} &=  \{ \Theta_{\min}, \Theta_{\min} +\Delta \Theta, \ldots, \Theta_{\max } \} \\
\bar{r} &= 10^{ \{  \lg(R_{\min}), \lg(R_{\min})+\Delta R, \ldots,  \lg(R_{\max})   \} },
\end{align}
\end{subequations}
where $\Delta \Theta$ and $\Delta R$ are the sampling intervals on the $\theta$ axis and $r$ axis, respectively, and
\begin{equation}
\Delta \Theta=\frac{\Theta_{\text {max }}-\Theta_{\text {min }}}{N_{\Theta}},  ~~\Delta R=\frac{\lg(R_{\text {max }})-\lg(R_{\text {min }})}{N_R}. 
\end{equation}

By employing the aforementioned method for spatial sampling, we can obtain Cartesian codebook and Polar codebook, denoted as
$\mathbf{U}_{\mathrm{C}}$ and $\mathbf{U}_{\mathrm{P}}$, respectively, and
\begin{subequations}
\begin{align}
\mathbf{U}_{\mathrm{C}}&=\left[\mathbf{u_c}\left(\bar{z}_0, \bar{x}_0\right), \mathbf{u_c}\left(\bar{z}_1, \bar{x}_1\right), \ldots, \mathbf{u_c}\left(\bar{z}_{N_Z N_X}, \bar{x}_{N_Z N_X}\right)\right]^{\top}, \\ % \label{eq:uc} 
\mathbf{U}_{\mathrm{P}}&=\left[\mathbf{u_p}\left(\bar{r}_0, \bar{\theta}_0\right), \mathbf{u_p}\left(\bar{r}_1, \bar{\theta}_1\right), \ldots, \mathbf{u_p}\left(\bar{r}_{N_R N_{\Theta}}, \bar{\theta}_{N_R N_{\Theta}}\right)\right]^\top. %\label{eq:up} 
\end{align}
\end{subequations}
Here, $\mathbf{u_c}\left(\bar{z}_i, \bar{x}_i\right) \in \mathbb{C}^{N \times 1}$ and $ \mathbf{u_p}\left(\bar{r}_i, \bar{\theta}_i\right) \in \mathbb{C}^{N \times 1}$ are the codewords {\bl and the $n$-th elements of them can be expressed as  
\begin{subequations}
\begin{align}
& {\left[\mathbf{u_c}\left(\bar{z}_i, \bar{x}_i\right)\right]_n=e^{j k_c d_n\left(\bar{z}_i, \bar{x}_i\right)}}, \\
& {\left[\mathbf{u_p}\left(\bar{z}_i, \bar{x}_i\right)\right]_n=e^{j k_c d_n\left(\bar{x}_i, \bar{\theta}_i\right)}}.
\end{align}
\end{subequations}}
And  
\begin{subequations}
\begin{align}
 d_n\left(\bar{z}_i, \bar{x}_i\right) &=\sqrt{\bar{z}_i^2+\left(\bar{x}_i^2-n \cdot \frac{d}{2}\right)^2},\\
 d_n\left(\bar{r}_i, \bar{\theta}_i\right) &=\sqrt{ \left(\bar{r}_i \cos(\bar{\theta}_i) \right)^2 + \left(\bar{r}_i \sin(\bar{\theta}_i)-n \cdot \frac{d}{2} \right)^2},
\end{align}
\end{subequations}
stand for the distance between the user or scatterer and the $n$-th antenna of the $i$-th codeword in the Cartesian coordinate system and the Polar coordinate system, respectively.

\begin{figure*}[t]
\centering
\includegraphics[height=3.75cm,width=18cm]{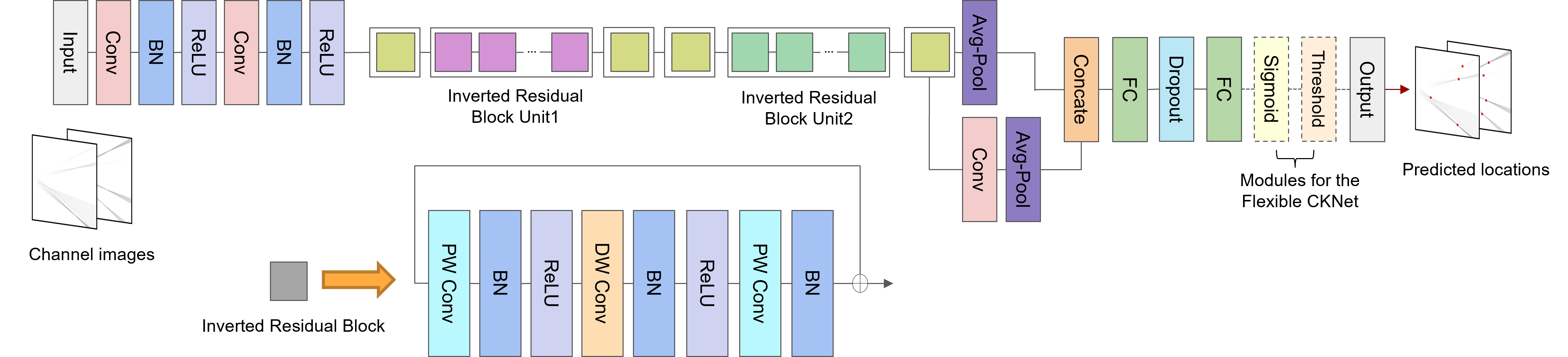}
\caption{The architectures of our proposed CKNet and FlexibleCKNet. CKNet is composed of some CBs, several IRBs, and IRBUs, the predicted coordinates are obtained by passing through two fully connected (FC) layers. Additionally, FlexibleCKNet adds a Sigmoid layer and a score filter at the end. The confidence scores for predicted keypoints are introduced, and a score filter is applied to remove the predictions with low scores.}
\label{cknet}
\end{figure*}

{\bl 
We reshape ${\mathbf{y}}_{\mathrm{T}}$ into a matrix ${\mathbf{Y}}_{\mathrm{T}} \in \mathbb{C}^{N_{T1} \times N_{T2}}$, where $N_{T1}$ and $N_{T2}$ represent the numbers of sampling points in the transformed domain. Next, we normalize the amplitude of each entry and transform the matrix into a grayscale $N_{T1} \times N_{T2}$ channel image by
\begin{equation}  % Comment: 為何用 || Y || ?，分子和分木不同? 注意 matrix norm 的定義 % Solved
{\mathbf{{Y}}}^\text{img}_{\mathrm{T}}={\left( 1-\frac{\lVert{\mathbf{{Y}}}_{\mathrm{T}}\rVert}{\max(\lVert{\mathbf{{Y}}}_{\mathrm{T}}\rVert)}
\right)}\times 255.
\label{eq:channelimage}
\end{equation}}
Each entry of ${\mathbf{{Y}}}^\text{img}_{\mathrm{T}}$ represents the pixel value ranging from 0 to 255, which indicates the grayscale level. And pixel value 0 corresponds to black and 255 represents white.

As shown in Fig. \ref{channelimages}(a), in the Cartesian domain, the channel image comprises numerous intersecting lines, where each pair of intersecting lines delineates a propagation path, with the energy being strongest at the intersection point and extending into the X-shaped energy convergence zones. Similarly, as shown in Fig. \ref{channelimages}(b), in the Polar domain, there are many intersecting curves forming hourglass-shaped energy convergence zones. Each of these regions represents a propagation path, with the energy being strongest at the intersection points. Therefore, we can regard these intersection points as keypoints and design the CKNet to detect these keypoints from channel images.

\subsection{CKNet}
We use the idea of grid cell-based object detection algorithms to divide the channel image into $S$ rows, and each row predicts 2 values to represent the position of a keypoint, i.e., $({z}_s^{\text {i }}, {x}_s^{\text {i}})$ or $({r}_s^{\text {i }}, {\theta}_s^{\text {i }})$. Building upon the positional features embedded in the transformed domain channel image, we design a keypoint detection network, namely CKNet. As depicted in Fig. \ref{cknet}, it is a lightweight CNN made up of various modules for precise and efficient detection of these intersections. We draw inspiration from MobileNetV2 \cite{mobilenetv2}, a lightweight model renowned for its superior performance in a multitude of computer vision tasks, such as object detection and segmentation. The architecture of CKNet is constructed leveraging the inverted residual block, a fundamental constituent of MobileNetV2. To capture features from the energy convergence zones of different sizes, we also utilize a multi-scale feature fusion mechanism. Details of each module are explained as follows.

\begin{table}[t]
\centering
\caption{Details of our CKNet, including the input size of each layer, the expansion ratio $t$ of the IRB and IRBU, the output channel size $c$, and the repetition count $n$.}
\label{tab1}
\begin{tabular}{ccccc}
   \toprule
   Input Size & Operator & t & c & n \\
   \midrule
   $1\times512\times512$ & CB & - & 64 & 1 \\
   $64\times256\times256$ & CB & - & 64 & 1 \\
   $64\times256\times256$ & IRB & 2 & 64 & 1 \\
   $64\times128\times128$ & IRBU & 2 & 64 & 4 \\
   $128\times64\times64$ & IRB & 2 & 128 & 1 \\
   $128\times64\times64$ & IRB & 4 & 128 & 1 \\
   $128\times64\times64$ & IRBU & 4 & 64 & 6 \\
   $128\times64\times64$ & IRB & 2 & 128 & 1 \\
   $16\times4\times4$ & Avg Pool & - & 16 & 1 \\
   $16\times64\times64$ & CB & - & 16 & 1 \\
   $16\times64\times64$ & CB & - & 32 & 1 \\
   $32\times32\times32$ & Avg Pool & - & 32 & 1 \\
   $1\times768$ & FC & - & 256 & 1 \\
   $1\times256$ & FC & - & 8 & 1 \\
   \bottomrule
\end{tabular}
\label{detailed_network}
\end{table}

{\bf (1) Composite Modules}:
\begin{itemize}
 \item Convolutional block (CB) consists of one convolutional layer, one batch normalization layer (BN), and one ReLU activation function layer.
 
\item Inverted Residual Block (IRB) is made up of one depth-wise convolutional layer (DW Conv) whose kernel size is $3 \times 3$, one BN layer, one ReLU activation function layer, one point-wise convolutional layer (PW Conv), one BN layer, one ReLU activation function layer, one DW Conv, one BN layer, and a skip connection from the input to the output of the last DW Conv.

\item Inverted Residual Block unit (IRBU) is made up of several IRBs.

\end{itemize}

{\bf (2) Network Modules}:

\begin{itemize}
\item Input layer:
The sparse channel images in transformed domains serve as the inputs.

\item Backbone:
The backbone is composed of several IRBs and IRBUs.

\item Multi-scale feature fusion:
We integrate feature maps from two different scales using two consecutive modules to extract intersection points from feature maps generated by the backbone. One module is an average pooling layer (Avg Pool), and the other is composed of a Conv layer, a BN layer, a ReLU activation layer, and an Avg Pool layer.

\item Output layer:
Finally, we use two FC layers to predict the coordinates of the user and scatterers. The output vector $ \hat{\mathbf{p}} \in \mathbb{R}^{1 \times 2{S}}$ represents ${S}$ output coordinates of the user and scatterers in the channel image, i.e., $\{(\hat{z}_s, \hat{x}_s)\}$ or $\{(\hat{r}_s, \hat{\theta}_s)\}, s = 1,\ldots,S$. 
\end{itemize}

We further transform the output coordinates of CKNet into the coordinates in Cartesian coordinate system or Polar coordinate system, i.e., $\{(\tilde{z}_s, \tilde{x}_s)\}$ or $\{(\tilde{r}_s, \tilde{\theta}_s)\}, s = 1,\ldots,S$. The transformation can be expressed as
\begin{subequations}
\begin{align}
&\left\{
\begin{aligned}
&\tilde{z}_s=\frac{\hat{z}_s}{I_W} \times\left(Z_{\text {max }}-Z_{\text {min }}\right)+Z_{\text {min}}, \\
&\tilde{x}_s=\frac{\hat{x}_s}{I_H} \times\left(X_{\text {max }}-X_{\text {min }}\right)+X_{\text {min}}.\\
\end{aligned}
\label{eq:coordinate_transform1}
\right.  \\
&\left\{
\begin{aligned}
&\tilde{\theta}_s=\frac{\hat{\theta}_s}{I_W} \times\left(\Theta_{\text {max }}-\Theta_{\text {min }}\right)+\Theta_{\text {min}}, \\
&\tilde{r}_s=\frac{\hat{r}_s}{I_H} \times \left(10^{\lg{R_{\text {max }}}}-10^{\lg{R_{\text {min }}}}\right)+10^{\lg{R_{\text {min}}}}.\\
\end{aligned}
\right. 
\label{eq:coordinate_transform2}
\end{align}
\end{subequations}

The detailed input and output dimensions of each layer are illustrated in Table \ref{detailed_network}. The IRB replace a full convolutional operator with a factorized version that splits convolution into two separate layers and greatly reduces the computational complexity. The first layer is a DW Conv, it performs lightweight filtering by applying a single convolutional filter per input channel. The second layer is a $1 \times 1$ convolution, namely a PW Conv, which is responsible for building new features through computing linear combinations of the input channels. This process effectively enriches the feature space, allowing the network to capture more complex features inherent in each feature layer. Therefore, CKNet can show good performance in detecting keypoints and has fast processing speed.
\begin{figure*}[t]
\centering
\includegraphics[height=2.8cm,width=15.26cm]{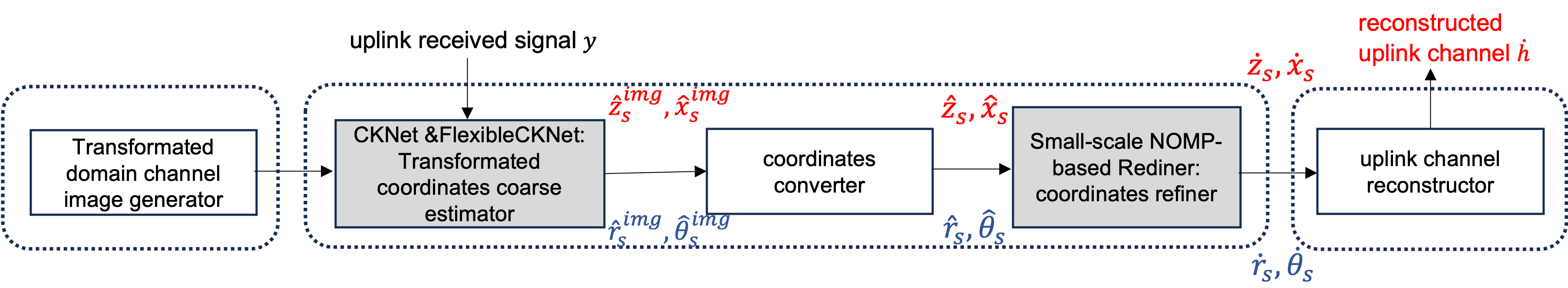}
\caption{Modules of the proposed channel reconstruction scheme, and we use red characters to represent the coordinates in the Cartesian domain and blue characters to represent the coordinates in the Polar domain.}
\label{algorithm}
\end{figure*}

{\bf (3) Loss function}:
We adopt the wing loss proposed in \cite{wingloss} to measure the distance between the predicted coordinates and the ground truth. It simultaneously possesses the advantages of both L1 and L2 loss functions. For small errors, it behaves as a logarithmic function with an offset, while for large errors, it behaves in an L1 pattern. This form of loss function enhances the capacity to handle errors within small to moderate ranges during training. Thus, it is well-suited for the keypoint detection task demanding high precision. Taking the Cartesian image as an example, for the $i$-th image, the outputs of CKNet are $\hat{\mathbf{p}}^i = (\hat{z}_s^i, \hat{x}_s^i)$, and the labels are $\mathbf{p}^i =(z_s^i, x_s^i)$. The loss function can be expressed as:
\begin{equation}  
\resizebox{\linewidth}{!}{
$
L_w (\mathbf{p}^i,\hat{\mathbf{p}}^i)=\left\{\begin{array}{ll}
w \ln (1+\|\mathbf{p}^i-\hat{\mathbf{p}}^i\| / \epsilon), & \text { if }\|\mathbf{p}^i-\hat{\mathbf{p}}^i\|<w, \\
\|\mathbf{p}^i-\hat{\mathbf{p}}^i\|-C, & \text { otherwise },
\end{array}\right. \text {, }$}
\end{equation}
where $w$ and $\epsilon$ are two hyper-parameters. $w$ is a positive number used to confine the nonlinear region within the interval $[-w, w]$, while $\epsilon$ controls the curvature, and $C$ is a small value for preventing gradient explosion.

\subsection{FlexibleCKNet}

In practical communication systems, the number of scatterers is commonly not fixed, resulting in an unpredictable number of propagation paths. Considering this situation, we raise the second question.

{\bf Q2: How can the keypoint detection network still be effective for extracting a variable number of keypoints?} Compared to the fixed number of keypoints detection that only predicts the location of keypoints, we additionally assign a confidence score to each keypoint representing the probability of the existence of each predicted keypoint. Assuming that the maximum possible number of paths is $S_{\max}$. The outputs are $({\hat{z}}_s^{\text {i }}, {\hat{x}}_s^{\text {i}}, \hat{C}_s^{\text {i }})$ or $({\hat{r}}_s^{\text {i }}, {\hat{\theta}}_s^{\text {i }}, \hat{C}_s^{\text {i }}), s = 1,\ldots,S_{\max}$, and $\hat{C}_s^{\text {i }}$ is the confidence score. Additionally, we use the sigmoid function to process the confidence scores, ensuring they are bounded between 0 and 1. The proximity of the score to 0 is inversely correlated with the probability of the path's existence. The architecture of FlexibleCKNet is depicted in Fig. \ref{cknet}, with the additional components compared to CKNet enclosed within dashed lines. We adjust the number of output nodes of the last FC layer from $S\times2$ to $S_{\max}\times3$.

{\bl 
The loss function includes not only the distance loss of the keypoint coordinates but also the evaluation of the predicted confidence score loss. It consists of two parts: one constrains the regression of coordinates using wing loss, while the other constrains the confidence scores using the binary cross-entropy function. The loss function can be expressed as
\begin{equation} 
\begin{aligned}
 Loss = & \frac{1}{M S_{\max}}  \sum_{i=1}^M \sum_{s=1}^{S_{\max}}  \mathbb{P}_{i,s}^{\mathrm{keypoint}} \lambda_{\rm coord}  L_w\left(p_s^i, \hat{p}_s^i\right) \\
& -C_s^i\log \left(\hat{C}_s^i\right)+\left(1-C_s^i\right) \log \left(1-\hat{C}_s^i\right),
\end{aligned}
\end{equation}
where $\mathbb{P}_{i,s}^{\mathrm{keypoint}}=0$ or $1$, with a value of 1 indicating that a keypoint exists in the $s$-th row of the $i$-th image, and $\hat{C}_s^i$ is the predicted score by FlexibleCKNet. The label of the confidence score is set as follows:

\begin{equation}
\begin{aligned}
& C_s^i=2\mathbb{P}_{i,s}^{\mathrm{keypoint}} \left(1-\text{sigmoid}\left(\frac{d}{\kappa
}\right)\right),
\end{aligned}
\end{equation}
where $\text {sigmoid}(x)=1/(1+e^{-x}) \in(0,1)$. By this setup, we constrain the ground truth of confidence score within $(0, 1)$. Here, $d$ represents the L1 distance between the predicted coordinates and the true coordinates. The smaller the distance, the closer the confidence score label is to 1; the larger the distance, the closer the confidence score label is to 0. $\kappa$ is a hyper-parameter that controls the convergence of the binary cross-entropy loss function. Additionally, another hyper-parameter, denoted as $\lambda_{\rm coord}$, is used to balance the two parts of the loss function.}

During the testing phase, we use a confidence threshold parameter, denoted as $\tau$, to filter out the predicted keypoints with low confidence scores. Therefore, we can detect all propagation paths with FlexibleCKNet and tackle the aforementioned question {\bf Q2}.

\section{Channel Reconstruction Scheme}

Drawing upon the channel model and leveraging the capabilities of our designed CKNet and FlexibleCKNet, we propose an efficient deep learning-based approach for the uplink channel reconstruction. Fig. \ref{algorithm} illustrates the schematic diagram of our proposed scheme, which operates sequentially through five successive modules.

$\bullet$ Module 1: Channel image generator. We encapsulate 
(\ref{eq:transform}-\ref{eq:channelimage}) into a channel image generator. By inputting the received antenna domain signal $\mathbf{y}$, we obtain the Cartesian domain or Polar domain channel image $\mathbf{{Y}}^\text{img}_{\mathrm{T}}$.

$\bullet$ Module 2: Keypoint detector. We utilize the previously designed CKNet or FlexibleCKNet as keypoint detectors to detect keypoint from the channel image and obtain the coordinates vector $\hat{\mathbf{p}}=\{(\hat{z}_s, \hat{x}_s)\}$ or $\hat{\mathbf{p}}=\{(\hat{r}_s, \hat{\theta}_s)\}, s = 1,\ldots,S$. We then use (\ref{eq:coordinate_transform1}) or (\ref{eq:coordinate_transform1})  to convert them into coordinates in the Cartesian domain or Polar domain, i.e., $\tilde{\mathbf{p}}=\{(\tilde{z}_s, \tilde{x}_s)\}$ or $\tilde{\mathbf{p}}=\{(\tilde{r}_s, \tilde{\theta}_s)\}, s = 1,\ldots,S$.

$\bullet$ Module 3: Small-scale NOMP Refiner. We employ a small-range codebook search and newtonized optimizer to refine the coarsely estimated parameters of each path. This process allows us to obtain more precise parameters represented as $\mathbf{\dot{p}}=\{(\dot{z}_s, \dot{x}_s)\}$ or $\mathbf{\dot{p}}=\{(\dot{r}_s, \dot{\theta}_s)\}$ and the complex gains $\mathbf{\dot{g}}=\{\dot{g}_s\}, s = 1,\ldots,S$.

$\bullet$ Module 4: Channel reconstructor. By substituting the estimated parameters $\mathbf{\dot{p}}$ and $\mathbf{\dot{g}}$ into (\ref{eq:channelmodel1}) or (\ref{eq:channelmodel2}), the channel can be finally reconstructed.

The proposed channel reconstruction scheme has relatively low computational complexity. In contrast to NNOMP algorithm, our approach can identify the parameters of all paths in a single-round network inference and reduce a significant amount of computational overhead. Modules 1-2 have been extensively discussed in the previous section. Detailed descriptions of module 3–4 are provided in the following subsections, along with further discussion on the scenario of flexible paths.

\subsubsection{Details of Module 3: Small-scale NOMP Refiner}

 The coarse estimation of channel parameters has been greatly accelerated by our well-trained CKNet instead of the exhaustive search on the whole codebook during each iteration when detecting a new path. While the neural network has achieved a relatively high detection accuracy at the level of input image resolution, the limitations of the input image resolution still leave room for refinement in estimating the positions of users or scatterers. Therefore, we employ a small-scale codebook to search around the positions estimated by the CKNet in a small scale and use newtonized optimizer to further improve the accuracy. The algorithm flow is illustrated in Algorithm 1.

\begin{figure}[t]
\centering
\includegraphics[height=4.2cm,width=8.6625cm]{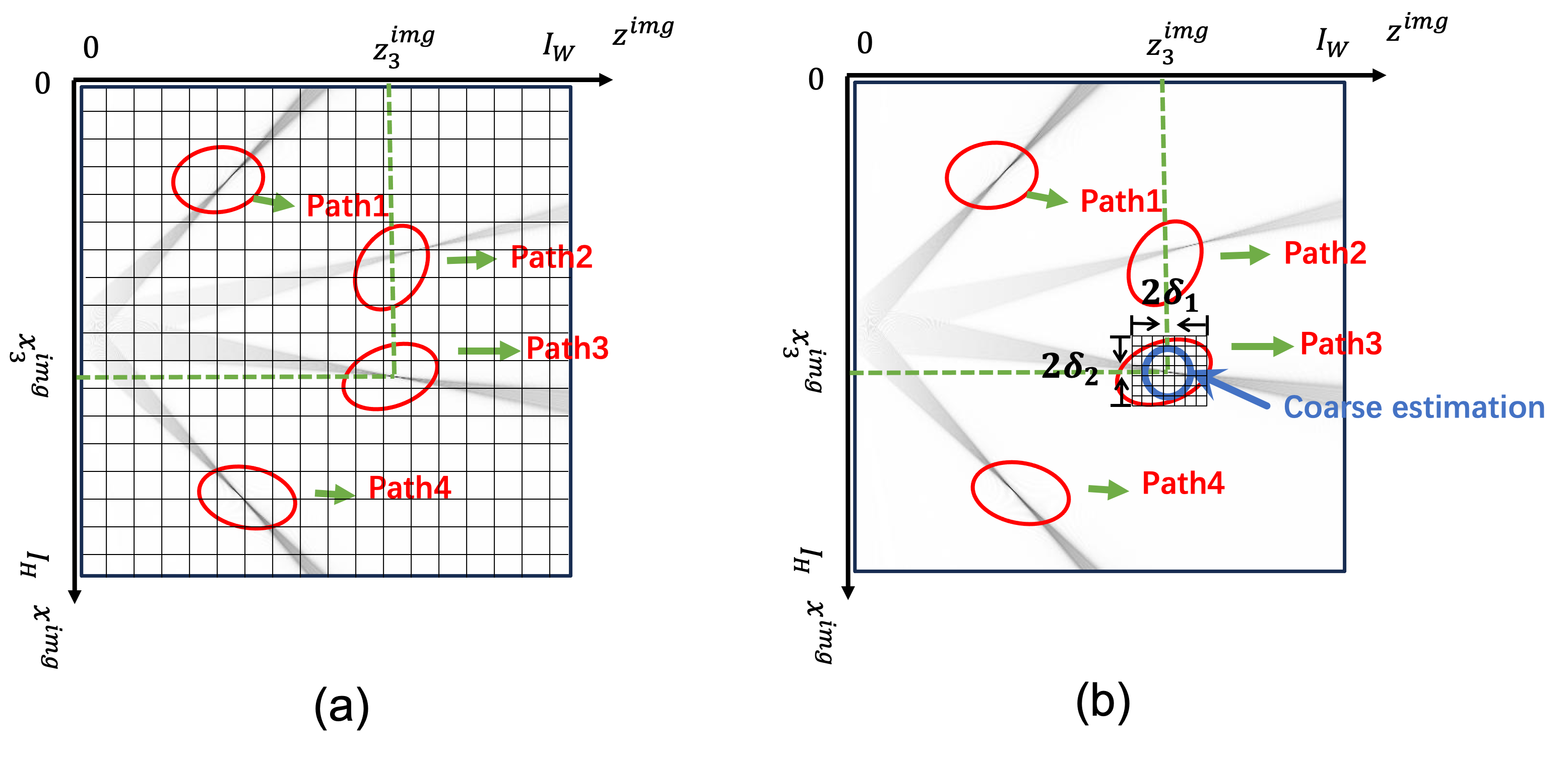}
\caption{
(a) Exhaustive search on the large codebook covering the whole observed region. (b) Fine-tune the more fine-grained and smaller codebook around the region surrounding the estimated coordinates.}
\label{codebooks}
\end{figure}

Following the method of NOMP, and given that we have already detected all paths, we begin with sorting all paths {\bl in ascending order} of the correlation coefficient between codeword and the residual signal, which can be expressed as

% \begin{equation}
% \mathbf{r}(s)=\frac{\left|\mathbf{a}\left(\tilde{p}_s\right)^{\top} \mathbf{y_r}\right|}{\left\|\mathbf{a}\left(\tilde{p}_s\right)\right\|_2}.
% \label{eq:correlation coefficient}
% \end{equation}

{\bl

\begin{equation}
\mathbf{r}(s)=\frac{\left\|\mathbf{a}\left(\tilde{p}_s\right)^{\top} \mathbf{y_r}\right\|}{\left\|\mathbf{a}\left(\tilde{p}_s\right)\right\|}
\label{eq:correlation coefficient}
\end{equation}
}

Sequentially, we conduct a fine-grained grid search within a $(2\delta1 \times 2\delta2 )$ region surrounding the estimated coordinates. We determine the searched coordinates as the position of the codeword having the maximum correlation coefficient with the residual signal $\mathbf{y_r}$ as depicted in (\ref{eq:small-scale omp}). Then, we conduct $R_s$ rounds of newtonized refinement.

% \begin{equation}
% \dot{p}_s = \max _{i, j} \frac{\left|\mathbf{U}_{T, i, j}^{\top} \mathbf{y}_r\right|}{\left\|\mathbf{U}_{T, i j}^{\top}\right)\right\|_2}.
% \label{eq:small-scale omp}
% \end{equation}
{\bl
\begin{equation}
\dot{p}_s=\max _{i, j} \frac{\left\|\mathbf{U}_{T, i, j}^{\top}\mathbf{y}_r\right\|}{\left\|\mathbf{U}_{T, i, j}^{\top}\right\|}
\label{eq:small-scale omp}
\end{equation}
}

After optimizing one path, we remove it from the estimation set $\tilde{\mathbf{p}}$, along with its portion in the residual signal. After each optimization, the parameters of the current path are added to $\mathcal{\dot{R}}$. When the small-scale NOMP is completed, we will obtain the final parameter set of all paths, i.e., $\mathcal{\dot{R}} = \{(\dot{z}_s, \dot{x}_s, \dot{g}_s)\},s=1...\hat{S}$ for the Cartesian domain, or $\mathcal{\dot{R}} = \{(\dot{r}_s, \dot{\theta}_s\, \dot{g}_s)\},s=1...\hat{S}$ for the Polar domain. 

\begin{algorithm}[t]
\caption{Small-scale NOMP Refiner}
\begin{algorithmic}[2]
\REQUIRE ~~ $\tilde{\mathbf{p}} = \{\tilde{p}_1,...,\tilde{p}_S\}$, $\mathcal{\dot{R}}=\{\}$\
\STATE \% sort all paths by the correlation coefficient in ascending order
\FOR{each $s=1:S$}
\STATE Calculate the steering vector $\mathbf{a}(\tilde{\mathbf{p}}_s)$ using (\ref{eq:steering_vector});
\STATE Calculate the correlation coefficient $\mathbf{r_s}$ using (\ref{eq:correlation coefficient});
\ENDFOR\

\STATE Sort correlation coefficient list $\mathbf{r}$;\
\STATE Reorder $\tilde{\mathbf{p}}$ into $\mathbf{\dot{p}}$;\

\FOR{each $s = 1:length(\mathbf{\dot{p}})$}
    \STATE Search in a small-scale codebook $(\dot{p}_s[0] \pm \delta1, \dot{p}_s[1] \pm \delta2)$ near the obtained point on the codebook and obtain the new location $\dot{p}_s$ with the highest correlation coefficient according to (\ref{eq:small-scale omp});\
    \STATE Update $\mathbf{\dot{p}}$;\
    \STATE Use Least Square algorithm to estimate $\mathbf{\dot{g}}$;\
    \STATE Do $R_s$ rounds of Newtonized single refinement, and obtain new $\dot{p}_s$ and $\mathbf{\dot{g}}$;\
    \STATE Add $\dot{p}_s$ to $\mathcal{\dot{R}}$;\
    \STATE Update the $\mathbf{\dot{g}}$ in $\mathcal{\dot{R}}$;\
    \STATE Remove $\dot{p}_s$ from $\mathbf{\dot{p}}$;\
    \STATE Use LS algorithm to estimate the current $\mathbf{\dot{g}}$;\
    \STATE Use (2) to calculate the steering vector $\mathbf{\dot{a}}$;\
    \STATE Update $\mathbf{y_r} = \mathbf{y_r} - \mathbf{\dot{g}} \mathbf{\dot{a}}(\mathbf{\dot{p}})$;\
\ENDFOR
\ENSURE ~~ $\mathcal{\dot{R}}$
\label{algorithm1}
\end{algorithmic}
\end{algorithm}

\subsubsection{Details of Module 4: Channel reconstructor}
Once we obtain the parameters $\mathbf{\dot{R}}$, the channel can be reconstructed by (\ref{eq:channelmodel1}) and (\ref{eq:channelmodel2}) for the Cartesian domain and the Polar domain, respectively.

\subsubsection{Channel reconstruction scheme under flexible-path scenario}

Our proposed CKNet-based channel reconstruction is designed based on the assumption of knowing the number of paths in advance. However, in real-world communication scenarios, the number of scatterers fluctuates based on environmental conditions. In such cases, efficiently performing user localization and channel reconstruction {\bl poses a challenge.} In the preceding section, we introduced the FlexibleCKNet, which can extract the coordinates of keypoints from channel images under scenarios with varying propagation path numbers. Due to the manually set confidence score threshold, such detection tasks may result in missed detections. To address this problem, we further incorporate a pre-judgment condition into the subsequent NOMP refiner. The detailed steps are described in Algorithm 2. When the residual energy exceeds a certain threshold $\tau_e$, an iterative search is conducted in the large-scale codebook to detect new path. After that, the {\bl small-scale} NOMP Refiner is performed. To be specific, the pre-judgment condition includes the following steps. First, we check the power of the residual signal
\begin{equation}
\mathbf{y}_r=\mathbf{y} - \sum_{s=1}^{\dot{S}} \dot{g}_s \mathbf{a}\left(\mathbf{\dot{p}}_s\right).
\end{equation}
When the residual power $\|\mathbf{y}_{r}\|^2$ is larger than the threshold $\tau_p=\sigma^2\sqrt{N} Q^{-1}\left(P_{f a}\right)+\sigma^2N$, where $P_{f a}$ is the false alarm rate, and $Q(x)=\int_x^{+\infty} 1/\sqrt{2 \pi} e^{-x^2/2} d x$ is the Gaussian Q function, the new detection step starts. We search over the whole codebook and obtain the new path with the highest correlation coefficient. And then, the small-scale NOMP refiner described in Table 1 begins to operate. Finally, we can obtain the fine-tuned estimations $(\dot{z}_s, \dot{x}_s, \dot{g}_s),s=1,\ldots,\dot{S}$. 

\begin{algorithm}[t]
\caption{Flexible Refiner}
\begin{algorithmic}[2]
\REQUIRE ~~ $\tilde{\mathbf{p}} = \{\tilde{p}_1,...,\tilde{p}_S\}$, $\mathbf{\dot{R}}={}$\
\STATE Calculate the complex gain $\mathbf{\tilde{g}}$ using LS algorithm;\
\STATE Calculate the residual power $\mathbf{y_r}=\mathbf{y} - \sum_{s=1}^{\tilde{S}} \mathbf{\tilde{g}} \mathbf{\tilde{a}}(\tilde{\mathbf{p}})$;\
\STATE \% Execute the decision criteria\
\WHILE{$\left\|\mathbf{y_r}\right\|^2 >= \tau_p$}
\STATE Search over the whole codebook to detect new path according to (\ref{eq:small-scale omp}).
\STATE Add $\tilde{p}_l$ to $\tilde{\mathbf{p}}$, $\tilde{S}=\tilde{S}+1$;\
\STATE Use LS algorithm to estimate the complex gain $\mathbf{\tilde{g}}$;\
\STATE Update $\mathbf{y_r} = \mathbf{y_r} - \mathbf{\tilde{g}} \mathbf{\tilde{a}(\tilde{\mathbf{p}})}$;\
\ENDWHILE\
\STATE Excute Algorithm 1.
\ENSURE ~~ $\mathbf{\dot{R}}$
\label{code:recentEnd}
\end{algorithmic}
\end{algorithm}

In most scenarios, the computational complexity remains manageable as long as we choose an appropriate confidence score threshold. The well-trained FlexibleCKNet can detect all paths effectively. Missed detections only occur in cases of minor path overlap, requiring a search across the entire codebook. Consequently, the additional computational complexity remains low on average. 
%%TBD

\section{EXPERIMENTAL RESULTS} 

In this section, we evaluate the performance of the proposed user localization and channel reconstruction scheme. Firstly, to visually demonstrate the effectiveness of our detection, we show some examples of our detection results under different SNR, various transformed domains, and in different scenarios where fixed paths and flexible paths. Then, to quantitatively evaluate the effectiveness of our proposed channel reconstruction scheme and the precision of user and scatterers localization, we utilize NMSE and L1 Distance as evaluation metrics. The calculation formulas are as follows:
\begin{equation}
\text { NMSE }=E\left\{\frac{\| \dot{\mathbf{h}}-\mathbf{h} \|^2}{\|\mathbf{h}\|^2}\right\},
\end{equation}

\begin{equation}
\text {L1} =\frac{1}{MS} \sum_{m=1}^M \sum_{s=1}^S\left|\dot{z}_s^m-z_s^m\right|+\left|\dot{x}_s^m-x_s^m\right|.
\end{equation}

\begin{figure}[t]
\centering
\includegraphics[height=6.5cm,width=7.98cm]{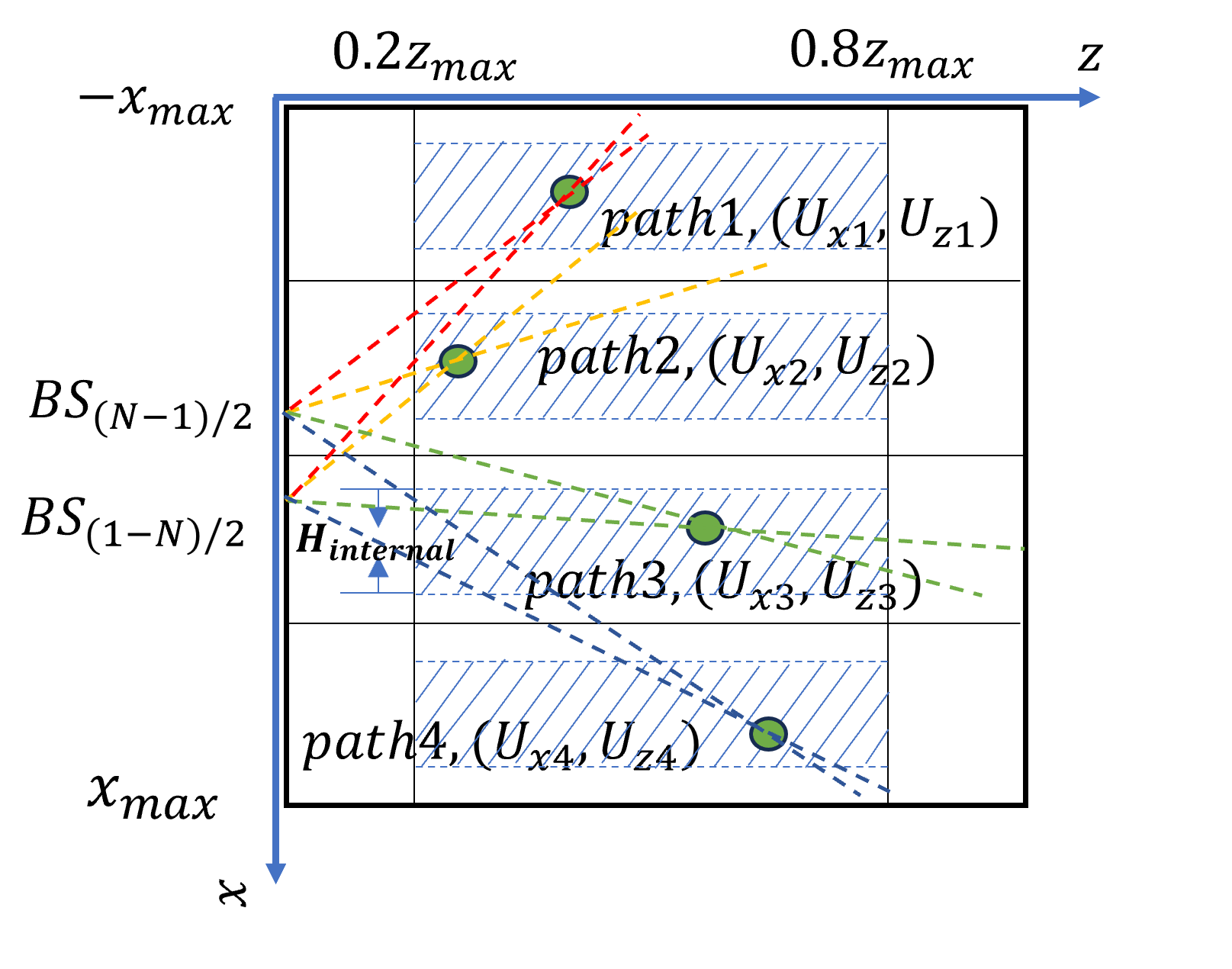}
\caption{The distribution of user and scatterers in the observed near-field region, and they are randomly distributed within evenly spaced areas.}
\label{distribution}
\end{figure}

\begin{figure*}[h]
\centering
\includegraphics[height=7.679cm,width=18cm]{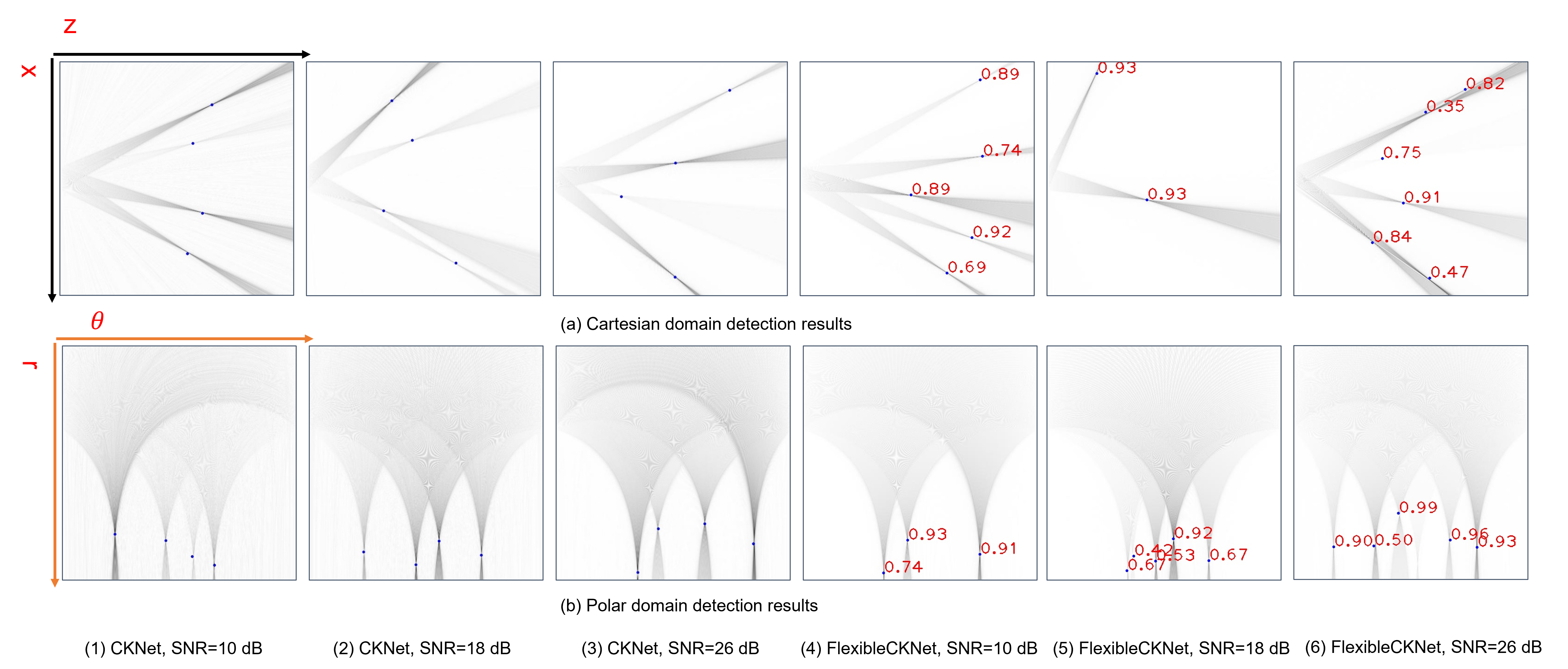}
\caption{The detection results of CKNet and FlexibleCKNet in the Cartesian domain and the Polar domain under different SNR scenarios. The detected user and scatterers are signed as blue spots on the channel images. The predicted confidence scores, marked in red, signify the predicted confidence regarding the existence of the user or scatterer.}
\label{show_results}
\end{figure*}

In the XL MIMO system, the uplink carrier frequency is $f_c=6$ GHz, the BS is equipped with $N=1024$ antennas. The number of path $S$ is set as 4 in the fixed-path scenario and the max number of path $S_{\max}$ is set as 6 in the flexible-path scenario, respectively. There is 1 user and $S-1$ scatterers. The spatial area of the observed region was defined by $[Z_{\min}, Z_{\max}]=[0, 5120\lambda]$ and $[X_{\min}, X_{\max}]=[-2560\lambda, 2560 \lambda]$, with sampling intervals of $10\lambda$. Correspondingly, we set $[\Theta_{\min}, \Theta_{\max}]=[-\pi/2, \pi/2]$, with sampling intervals of $0.002pi$, $[R_{\min}, R_{\max}]=[100\lambda, 5120\lambda]$. For the small-scale NOMP, we set $\delta1=\delta2=20\lambda$, and the sampling interval is $\lambda$.

% and the sampling interval of $lg(R)$ is $0.08pi$

To train and evaluate the CKNet and FlexibleCKNet, we generated 1800, 600, and 120 channel images in the Cartesian domain and Polar domain for the training, validation, and testing datasets, respectively, covering a range of SNR from $10$ dB to $26$ dB. The input sizes of both CKNet and FlexibleCKNet were set to $I_W \times I_H = 512 \times 512$. During the training phase, we employed the Adam optimizer with an initial learning rate of $2e-4$ and weight decay of $1e-4$, and trained for 2000 iterations, and the training and validation batch sizes are 16 and 8, respectively. For the hyper-parameters in loss function, we set $w=10$, $\epsilon=5$, and $\kappa=5$. Additionally, for the inference of FlexibleCKNet, we set the confidence score threshold $\tau=0.3$ and filter out predicted points with scores less than this value. Both of CKNet and FlexbleCKNet are applicable to  Cartesian domain datasets and Polar domain datasets, and we conducted separate training and testing.

 During generating our dataset, in the scenario with a fixed number of paths, the distribution of user and scatterers has a certain separation, following the following pattern illustrated as Fig. \ref{distribution}. {\bl The distribution area of users and scatterers is divided into S regions vertically. In other words, different users and scatterers are allocated to distinct angular intervals, and the height of each region is $H_\text{interval}$.} Within each region, user or scatterers are randomly distributed. Their positions adhere to the following formula:
 {\bl
\begin{equation}
\begin{aligned}
U_{x, s} &\in \Big[-X_{\max }   +\frac{2 X_{\max }}{S}(s-1)+\frac{1}{2}\left(\frac{1}{h_{\text {ratio }}}-1\right)  \cdot H_{\text {interval}}, \\
 &  -X_{\max }+\frac{2 X_{\max }}{S} \cdot s-\frac{1}{2}\left(\frac{1}{h_{\text {ratio }}}-1\right)  \cdot H_{\text {interval}} \Big], \\
U_{z, s} &\in\left[0.2 \cdot Z_{\max }, 0.8 \cdot Z_{\max }\right].
\end{aligned}
\end{equation}
}

\begin{figure}[t]
\centering
\includegraphics[height=6cm,width=8cm]{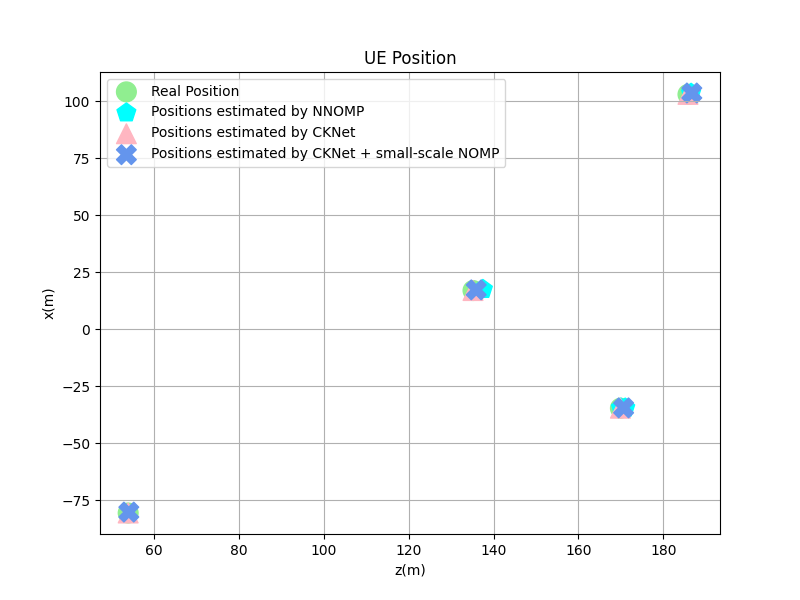}
\caption{The real and estimated locations of user and scatters by different algorithms.}
\label{UEpos}
\end{figure}
And we set $h_{\text {ratio }}=1/2$, and the intervals between different regions of user and scatterers are all set as $H_{\text {internal }}=2 \cdot X_{\max }/S \cdot h_{\text {ratio }}$. For the FlexibleCKNet, during the training phase, we set the weight of the regression loss $\lambda_{\rm coord}=10$ for the first 1000 iterations and $\lambda_{\rm coord}=1.0$ for the following 1000 iterations. The setting of regression loss prompts the network to initially focus on learning the confidence, before gradually adjusting the coordinate loss. This helps speed up the convergence of the network. Our experiments were conducted on the Windows 11 with a 12th Gen Intel(R) Core(TM) i7-12700 CPU and NVIDIA Tesla V100-SXM2 GPU.

\begin{figure*}[t]
\centering
\subfigure[]{
\includegraphics[scale=0.48]{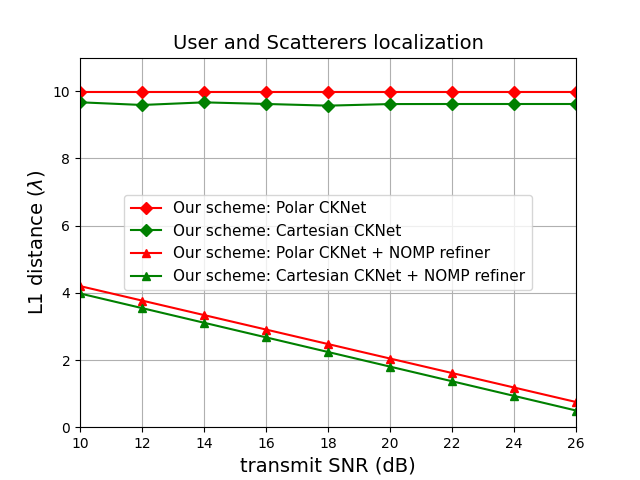}\label{numerical_results21}
}
\centering
\subfigure[]{
\includegraphics[scale=0.48]{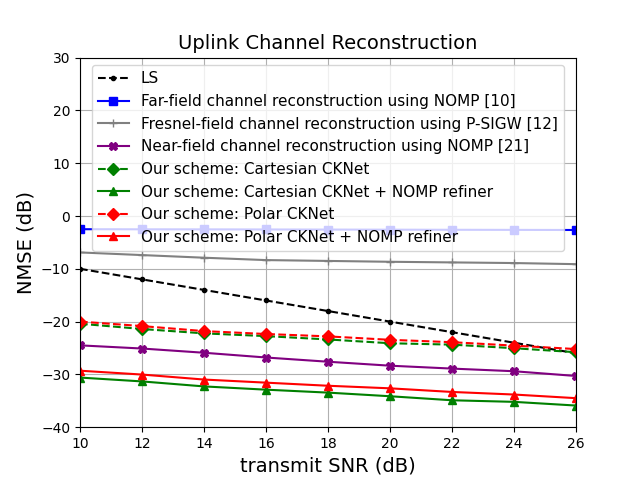}\label{numerical_results22}
}
\caption{CKNet: {\bl (a) User and scatterers localization performance.} We compare the L1 Distance of the outputs of CKNet under the scenarios with {\bl small-scale} NOMP refine and without {\bl small-scale} NOMP refine in Polar domain and Cartesian domain, respectively. {\bl (b) Channel reconstruction performance.} We compare the proposed channel reconstruction mechanism with LS estimation, Far-field NOMP \cite{5}, Fresnel-field P-SIGW \cite{8}, and Near-field NOMP \cite{nnomp1}.}
\label{numerical_results}
\end{figure*}

\subsection{Localization performance of CKNet and FlexibleCKNet}
\begin{figure*}[t]
\centering
\subfigure[]{
\includegraphics[scale=0.48]{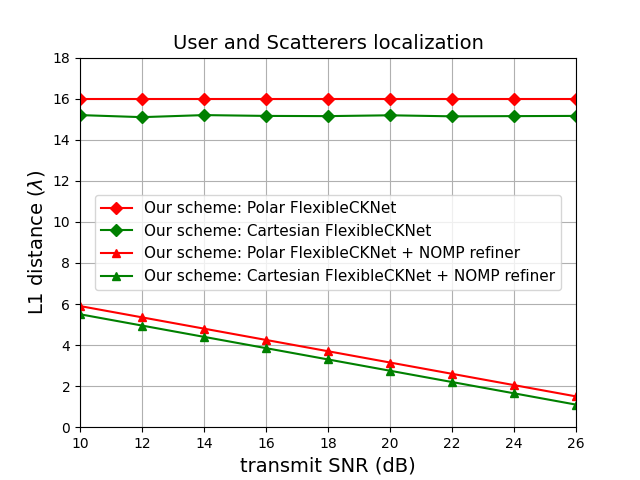}\label{numerical_results221}
}
\centering
\subfigure[]{
\includegraphics[scale=0.48]{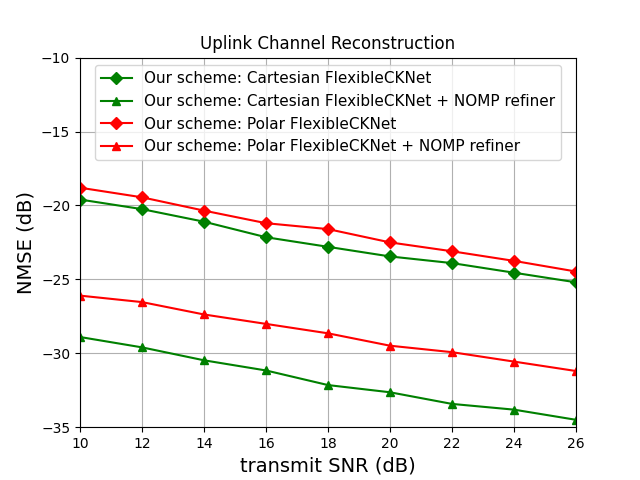}\label{numerical_results222}
}
\caption{FlexibleCKNet: {\bl(a) User and scatterers localization performance.} (b) Channel reconstruction performance.  We compare the performance of FlexibleCKNet under the scenarios with {\bl small-scale} NOMP refine and without {\bl small-scale} NOMP refine in Polar domain and Cartesian domain, respectively.}
\label{numerical_results2}
\end{figure*}

Fig. \ref{show_results} illustrates the detection results for the Catersian domain channel images and the Polar domain channel images under the scenario of various SNR. The first three columns are examples of the detection results of CKNet, and the last three columns are examples of detection results of FlexibleCKNet. It is noticeable that across different SNR, the intersection points within the channel images can be detected accurately. Our network exhibits excellent detection performance, effectively identifying even the paths with low energy in the channel images, as depicted in Fig. \ref{show_results} (a-1). Moreover, our CKNet and FlexibleCKNet demonstrate great versatility across the Cartesian domain and the Polar domain, effectively capturing the characteristics of intersection points in both types of channel images and providing a general detection network. 

{\bl Fig. \ref{UEpos} shows the real locations of users and scatterers, as well as the estimations by NNOMP, CKNet, and CKNet with a small-scale NOMP refiner. All of these algorithms can precisely locate users and scatterers.} We utilize the L1 distance for quantitative evaluation of localization accuracy and the results are depicted in Fig. \ref{numerical_results} (a). For the Cartesian domain channel images, the L1 detection error under different SNR is consistently around $9.7 \lambda$, and the actual value is approximately 0.485 meters. However, the detection error in the Polar domain is slightly higher, at around $10 \lambda$ (approximately 0.5 meters). This is because the coverage area of hourglass-shaped propagation paths in the Polar domain is larger than that of X-shaped propagation paths in the Cartesian domain, leading to a higher probability of user or scatterers falling into other path regions. Such a case where the path falling into overlapped areas can cause a decrease in detection accuracy as shown in Fig. \ref{show_results} (b-5). 

{\bl It can also be observed from the last three columns that the FlexibleCKNet not only predicts the coordinates of intersection keypoints but also assigns a confidence score to each keypoint. For the users or scatterers not located in the overlapped areas, they are more likely to be detected with high precision, and the confidence scores are also high. While for users or scatterers that fall into the overlapped areas, both confidence scores and detection accuracy will slightly decrease. For example, the paths of the first and the sixth scatterers in Fig. \ref{show_results} (a-6) have fallen into the energy area of other path and their confidence scores are 0.35 and 0.47, and both of them are lower than the other paths with similar energy. Additionally, when the confidence score is below 0.1, the path is considered nonexistent, resulting in missed detection (low recall), as shown in Fig. \ref{show_results} (b-6), the fourth path with low energy is filtered with a confidence score of 0.087. In such case, we apply the {\bl small-scale} NOMP refinement in the following phase and re-detect the missing paths. Under various SNR scenarios, our FlexibleCKNet can accurately detect all paths and is applicable to both Cartesian domain channel images and Polar domain channel images. The L1 Distance of FlexibleCKNet is shown in Fig. \ref{numerical_results2} (a), the L1 Distance remains around $15.2 \lambda$ and $15.9 \lambda$ under different SNR scenarios for the Cartesian domain and Polar domain channel images, respectively. The recall rate is 0.95, which shows the detection algorithm can identify most of the existing paths. The designed CKNet and Flexible CKNet with high recall rate and detection accuracy are good coarse estimators which serve as the foundation for high-precision, fast channel reconstruction and are crucial for the overall algorithm. }

% \begin{figure*}[t]
% \centering
% \includegraphics[height=6.25cm,width=15cm]{2.png}
% \caption{FlexibleCKNet: (a) User and scatterers localization performance. (b) Channel reconstruction performance.  We compare the performance of FlexibleCKNet under the scenarios with {\bl small-scale} NOMP refine and without {\bl small-scale} NOMP refine in Polar domain and Cartesian domain, respectively.}
% \label{numerical_results2}
% \end{figure*}

\subsection{Evaluation of the small-scale NOMP refinement} 

To evaluate the performance of the small-scale NOMP refinement in localization and channel reconstruction, we compare the L1 Distance of the predicted coordinates and the true locations and the NMSE of the reconstructed and original channel matrix.
As Fig. \ref{numerical_results} (a) shows, under the scenario of fixed propagation paths, after applying the small-scale NOMP refinement, there is a further noticeable decrease in localization error. Moreover, this decrease gradually grows with the increase in SNR, reaching an L1 distance of $0.5 \lambda$ around 26 dB. Similarly, as shown in Fig. \ref{numerical_results2} (a), in the flexible path scenario, the small-scale NOMP refinement further refined the target positions, improving the localization accuracy and reaching an L1 distance of $1.1 \lambda$ around 26 dB in the Cartesian domain.

\subsection{Comparison of the proposed channel reconstruction scheme and other algorithms} 
We further evaluate the proposed channel reconstruction scheme with the benchmarks of Far-field channel reconstruction with NOMP \cite{5}, Fresnel-field channel reconstruction with P-SIGW \cite{8}, and Near-field channel reconstruction with NOMP \cite{nnomp1}. They are all iterative codebook-based methods. Fig. \ref{numerical_results} (b) and Fig. \ref{numerical_results2} (b) present the NMSE performance of the reconstructed channels. It can be observed that directly using the far-field codebook yields the worst performance with an NMSE of approximately remains at -2.5 dB, demonstrating the distinct characteristics of near-field and far-field regions. The same for applying the PISGW algorithm with the Fresnel region codebook. The NNOMP scheme in \cite{nnomp1} leverages the sparsity of the near-field region in the Polar domain utilizing the NOMP algorithm for exhaustive search and achieves decent NMSE performance. In our keypoint-empowered channel reconstruction scheme, due to the limitations in the resolution of channel image, the keypoint positions obtained by CKNet or FlexibleCKNet may not be as precise as those obtained by NNOMP through large-scale codebook search. Therefore, the coarse estimated channel accuracy shows higher NMSE than NNOMP. However, our small-scale fine-grained NOMP refiner and newtonized optimizer help to further improve the performance. Our channel reconstruction scheme outperforms NNOMP by approximately 5 dB at different SNRs, demonstrating high channel estimation accuracy. As shown in Fig. \ref{numerical_results2}, in scenarios with a flexible number of paths under different SNR, our algorithm can also achieve performance close to that with a fixed number of paths.

\subsection{Analysis of the computational complexity}

{\bl
% We design the lightweight CKNet to extract keypoints from the channel image, replacing the exhaustive search over a large-scale codebook in NNOMP algorithm, thereby significantly reducing the computational complexity. In our experiments, the codebook used for NNOMP, with $N_{X1}$ rows and $N_{Z1}$ columns, is set to be twice the size of the codebook used for generating images in our algorithm ($N_{X2} \times N_{Z2}$) to achieve a relatively high detection accuracy. As a result, the proposed algorithm can achieve excellent localization and channel reconstruction performance while saving a considerable amount of computational cost. 

% Commnet: 直接用 N' 就可以，不需要 N^{'}
% Solved

Table \ref{complexity} compares the computational complexity of our proposed channel reconstruction mechanism and the NNOMP algorithm, detailing the complexity of each step. The dominant part is the coarse estimation step, while our algorithm includes an additional image generation step. $N_{X}', N_{X}$, $N_{Z}'$, and $N_{Z}$ represent the sampling numbers of the $x$-axis and $z$-axis in the Cartesian domain, respectively. $K$ denotes the kernel size of each convolutional layer, and $E$ represents the total number of features of the IRB.

In our experiments, the codebook used for NNOMP, with $N_{X}'$ rows and $N_{Z}'$ columns, is set to be twice the size of the codebook used for generating images in our algorithm ($N_{X} \times N_{Z}$) to achieve relatively high detection accuracy. Compared to the complexity of the coarse estimation in NNOMP, which is $\mathcal{O}(S N N_{X}' N_{Z}')$, the complexity of image generation, $\mathcal{O}(N N_{X} N_{Z})$, is relatively low.
For the coarse estimation, CKNet requires only a single forward propagation to obtain parameters of all paths, replacing the exhaustive search for all paths on a large-scale two-dimensional codebook. Additionally, the computation of CKNet involves only multiplication and addition operations, while NNOMP also includes the pseudo-inverse matrix operation, which is significantly more computationally complex to implement.
For the small-scale codebook search, the computational complexity is also much lower than that of searching over a large-scale codebook in NNOMP because $N_{\delta1} \ll N_{X}'$ and $N_{\delta2} \ll N_{Z}'$. The proposed channel reconstruction scheme provides a practical solution for real communication systems.

\begin{table}[t]
\centering
\caption{{\bl Comparison of the computational complexity.}} % 添加标题
\label{complexity} % 设置表格的标签
\begin{tabular}{lll} % l 表示左对齐
\toprule
 & {\bl NNOMP} & {\bl Our proposed algorithm}  \\ % 顶部横线
\midrule
{\bl Image Generation}   & {\bl -} & {\bl$\mathcal{O}(N N_{X} N_{Z})$} \\
{\bl Coarse Estimation}   & {\bl $\mathcal{O}(S N N_{X}' N_{Z}')$} & {\bl $\mathcal{O}\left(T K^2 E\right)$}  \\
{\bl Refinement}& {\bl $\mathcal{O}(R_{c}R_{s} S^2 N)$}    & {\bl $\mathcal{O}(S N_{\delta_1} N_{\delta_2})$  + $\mathcal{O}(R_{s} S^2 N)$} \\
\bottomrule % 底部横线
\end{tabular}
\end{table}

}

% The average running time of NNOMP is 33.72s, and the average running time of our channel reconstruction scheme is 2.39 s (image generation 2.18 s, channel reconstruction 0.21 s), which is reduced by 14 times and provides a solution for the application in real communication systems.

\section{Conclusion}
This paper considered the near-field region in the XL MIMO system and proposed a keypoint detection-empowered user localization and channel reconstruction scheme\textit{}. Two key problems on the computational complexity and the flexible path numbers in the real communication systems were successfully tackled by CKNet and FlexibleCKNet. An efficient user localization and channel reconstruction scheme transforming the received signal into channel image and designing CNNs to extract the user locations from the image. A channel reconstructor was proposed to improve the detection and channel estimation accuracy. The numerical results show the efficiency of the proposed user localization and channel reconstruction scheme. The user and scatters can be accurately located and the channel reconstruction accuracy is also superior to that of the iterative codebook-based schemes in the far-field region, Fresnel-field region, and near-field region, respectively. Additionally, our method achieves a reduction in computational complexity by orders of magnitude, showcasing its applicability in real communication systems.
% \section{ACKNOWLEDGEMENT}
\appendix{
\subsection{Proof of Property 1}
{\bl
Take the Cartesian domain transformation as an example, the $i$-th element of the transformed received signal is
\begin{equation}
\begin{aligned}
\left||Y_c|\right|_i & =\left||\mathbf{u}_{C, i} \cdot \mathbf{y}|\right| \\
% & =\left|\mathbf{c}\left(\bar{z}_i, \bar{x}_i\right) \cdot \mathbf{y}\right| \\
& =||\mathbf{c}\left(\bar{z}_i, \bar{x}_i\right) \cdot\left[\sum_{s=1}^S \sqrt{P} g_s \mathbf{a}(z_s, x_s)+\mathbf{n}\right]||.\\
\end{aligned}
\label{yci}
\end{equation}
The noise variance is much smaller than that of the transmitted signal. Therefore, the noise term can be ignored in (\ref{yci}). 
\begin{equation}
\resizebox{0.5\textwidth}{!}{$
\begin{aligned}
\left\| |Y_c| \right\|_i & = \left\| \sum_{s=1}^S \sqrt{P} g_s \cdot \left[ e^{-j k_c d_{-\frac{1-N}{2}}\left(\bar{z}_i, \bar{x}_i\right)}, \dots, e^{-j k_c d_{\frac{N-1}{2}}\left(\bar{z}_i, \bar{x}_i\right)} \right] \right. \\
& \quad \left. \cdot \left[ \frac{1}{d_{\frac{1-N}{2}}(z_s, x_s)} e^{j k_c d_{\frac{1-N}{2}}(z_s, x_s)}, \dots, \frac{1}{d_{\frac{N-1}{2}}(z_s, x_s)} e^{j k_c d_{\frac{N-1}{2}}(z_s, x_s)} \right]^T \right\| \\
& = \sum_{s=1}^S \sqrt{P} \left\| |g_s| \right\| \cdot \left\| \sum_{n=\frac{1-N}{2}}^{\frac{N-1}{2}} \frac{1}{d_n(z_s, x_s)} \cdot e^{-j k_c \left[ d_n\left(\bar{z}_i, \bar{x}_i\right) - d_n(z_s, x_s) \right]} \right\| \\
& \leq \sum_{s=1}^S \sqrt{P} \left\| |g_s| \right\| \cdot \sum_{n=\frac{1-N}{2}}^{\frac{N-1}{2}} \frac{1}{d_n(z_s, x_s)} \cdot \left\| e^{-j k_c \left[ d_n\left(\bar{z}_i, \bar{x}_i\right) - d_n(z_s, x_s) \right]} \right\| \\
& = \sum_{s=1}^S \sqrt{P} \left\| |g_s| \right\| \cdot \frac{1}{d_n(z_s, x_s)} \cdot N.
\end{aligned}$
}
\label{prove1}
\end{equation}

(\ref{prove1}) holds if only if for all $n$, $d_n\left(\bar{z}_i, \bar{x}_i\right)-d_n(z_s, x_s)$ takes the same value. That is, $\bar{z}_i=z_{s,}, \bar{x}_i=x_s$. At this value, $\left|Y_c\right|_i$  attains its maximum. The same principle applies to the Polar domain.

 \small\bibliographystyle{IEEEtran}
 \bibliography{main}

\vspace{12pt}
\color{red}

\end{document}